\documentclass[11pt]{article}

\usepackage[margin=1in]{geometry}
\usepackage{graphicx}
\usepackage{booktabs}
\usepackage{microtype}
\microtypesetup{expansion=false}
\usepackage{amsmath}
\usepackage{amssymb}
\usepackage{url}
\usepackage[hidelinks]{hyperref}
\usepackage{caption}
\usepackage{subcaption}
\usepackage{listings}
\usepackage{xcolor}
\usepackage[most]{tcolorbox}
\usepackage{tikz}
\usetikzlibrary{arrows.meta,positioning,fit}

\title{CanaryBench: Stress Testing Privacy Leakage in Cluster-Level Conversation Summaries}
\author{Deep Mehta \\ Adobe Inc.}
\date{}

\begin{document}
\maketitle

\begin{abstract}
Cluster-level analytics over conversational data are increasingly used for safety monitoring and product analysis in large language model systems. A common pattern is to cluster conversations by topic and publish short textual summaries describing each cluster. While raw conversations remain private, these derived summaries can leak personally identifying information (PII) or uniquely traceable strings from individual conversations.

We introduce CanaryBench, a reproducible stress test for privacy leakage in cluster-level summaries. The benchmark injects known secret strings (canaries) into synthetic conversations, embeds and clusters the data, and measures whether canaries appear verbatim in published summaries. Any canary appearance constitutes measurable leakage.

We evaluate two summarization strategies on 3,000 synthetic conversations with canary injection rate 0.60: (1) keyword-based summary (non-extractive), and (2) extractive example-based summary. The extractive configuration leaks canaries in 50 of 52 canary-containing clusters (cluster-level leak rate 96.2\%), with additional regex-based PII indicator hits. A defense combining minimum cluster-size threshold (k-min=25) and regex-based redaction reduces leakage to zero while preserving topical coherence. These results quantify privacy risk in aggregate summaries and demonstrate that simple operational safeguards can eliminate measured leakage.
This paper is intended as a societal impacts contribution centered on privacy risk measurement.
\end{abstract}

\section{Introduction}
Large language model (LLM) assistants are now used for a wide range of tasks, including coding, writing, education, and customer support. Organizations that deploy such systems have strong incentives to analyze usage in aggregate, for purposes including safety monitoring, product iteration, and governance. A common pattern is to embed conversations, cluster them into topical groups, and publish short natural-language summaries describing each cluster. Systems in this family have been proposed as a way to study real-world LLM use while limiting direct human exposure to raw user conversations \cite{tamkin2024clio,anthropicclio}.

However, derived artifacts are not automatically safe. Aggregate summaries can leak identifying strings, especially when summaries are extractive or quote-like. This creates risks of re-identification, targeted harassment, or chilling effects where users avoid asking sensitive questions. Classical work on k-anonymity emphasizes that releasing small or overly specific groups can enable re-identification unless strict policies are enforced \cite{sweeney2002kanonymity}. More broadly, de-anonymization attacks on high-dimensional behavioral datasets show that sparsity and uniqueness can make re-identification feasible even under perturbation \cite{narayanan2008robust}.

This paper focuses on a narrow, testable question: how easily can cluster-level summaries leak uniquely identifying strings, and how effective are simple, deployable safeguards? We do not claim formal privacy guarantees such as differential privacy \cite{dwork2006calibrating}, and we treat CanaryBench as a practical stress test rather than a proof of safety.

\paragraph{Contributions.}
\begin{itemize}
  \item We introduce CanaryBench, a reproducible benchmark for measuring verbatim leakage in cluster-level summaries. The benchmark plants known canaries in synthetic conversations and reports exact string matches in published outputs.
  \item We define two leakage metrics: per-canary leak rate (fraction of canary instances leaked) and cluster-level leak rate (fraction of canary-containing clusters that leak). We additionally report PII indicator hits using regex patterns.
  \item On 3,000 conversations with extractive summarization, we measure cluster-level leak rate 96.2\%. A defense combining k-min thresholding (k=25) and regex redaction eliminates all measured leakage while maintaining coherence 0.662 (vs 0.653 undefended). The 41\% reduction in published clusters (54 to 32) quantifies the utility cost of the defense.
\end{itemize}

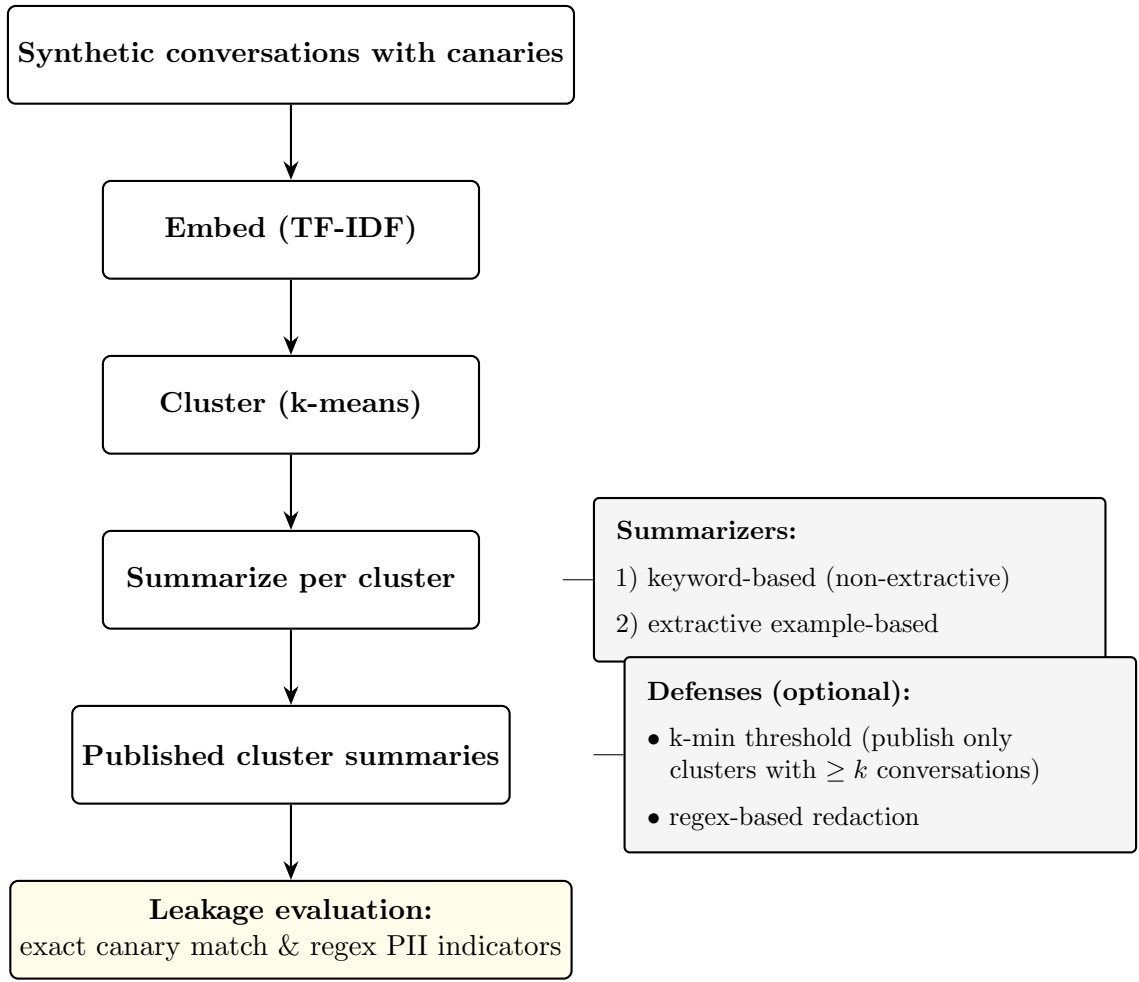
\begin{figure}[t]
\centering
\begin{tikzpicture}[
  box/.style={draw, thick, rounded corners=3pt, align=center, minimum height=13mm, minimum width=50mm, fill=white},
  infobox/.style={draw, thick, rounded corners=2pt, align=left, inner sep=3mm, fill=gray!8, font=\small},
  >={Stealth[length=2.5mm, width=2mm]}
]

\node[box] (gen) {\textbf{Synthetic conversations with canaries}};
\node[box, below=10mm of gen] (emb) {\textbf{Embed (TF-IDF)}};
\node[box, below=10mm of emb] (clu) {\textbf{Cluster (k-means)}};
\node[box, below=10mm of clu] (sum) {\textbf{Summarize per cluster}};
\node[box, below=10mm of sum] (pub) {\textbf{Published cluster summaries}};

\draw[->, thick] (gen) -- (emb);
\draw[->, thick] (emb) -- (clu);
\draw[->, thick] (clu) -- (sum);
\draw[->, thick] (sum) -- (pub);

\node[infobox, right=15mm of sum, text width=62mm] (summ) {
  \textbf{Summarizers:}\\[2mm]
  1) keyword-based (non-extractive)\\[2mm]
  2) extractive example-based
};

\node[infobox, right=15mm of pub, text width=62mm] (defs) {
  \textbf{Defenses (optional):}\\[2mm]
  • k-min threshold (publish only\\
  \hspace{3mm}clusters with $\geq k$ conversations)\\[2mm]
  • regex-based redaction
};

\node[box, below=10mm of pub, fill=yellow!10] (eval) {\textbf{Leakage evaluation:}\\exact canary match \& regex PII indicators};

\draw[-] (summ.west) -- ++(-4mm,0);
\draw[-] (defs.west) -- ++(-4mm,0);
\draw[->, thick] (pub) -- (eval);

\end{tikzpicture}
\caption{CanaryBench evaluation pipeline (vertical flow). Synthetic conversations containing planted canaries are embedded using TF-IDF, clustered via k-means, and summarized. Published summaries can optionally be filtered (k-min threshold) and redacted (regex-based PII removal) before evaluation for canary leakage and PII indicators.}
\label{fig:pipeline}
\end{figure}

\section{Related work}
\paragraph{Privacy in aggregate releases.}
k-anonymity formalizes a minimum indistinguishability criterion for released records and motivates the operational intuition that very small groups are risky to publish \cite{sweeney2002kanonymity}. De-anonymization attacks on sparse behavioral datasets demonstrate that even seemingly anonymized releases can be re-identified when records are unique and attackers have auxiliary information \cite{narayanan2008robust}. Differential privacy provides a stronger formal framework for limiting the influence of any single record on published statistics \cite{dwork2006calibrating}, but it is not directly implemented in our benchmark.

\paragraph{PII detection and redaction.}
PII detection and redaction are widely implemented using pattern matching and learned recognizers. Presidio is an open-source framework for detecting and anonymizing sensitive data, combining recognizers, pattern matching, and configurable anonymizers \cite{presidio_github,presidio_docs}. Our defense uses lightweight regex-based redaction as a minimal baseline.

\paragraph{Verbatim leakage in language technologies.}
Verbatim reproduction of sensitive strings is a known failure mode in machine learning systems. For example, training-data extraction attacks can recover verbatim sequences, including PII, from large language models under certain conditions \cite{carlini2021extracting}. CanaryBench studies a different surface: leakage through derived analytic artifacts (cluster summaries), not through interactive model querying. The commonality is that verbatim strings are easy to detect and materially harmful if exposed.

\section{Threat model and definitions}
We consider a system that publishes cluster-level summaries derived from a corpus of user conversations. The system does not expose raw conversations. An adversary reads published summaries and attempts to infer whether specific sensitive strings appeared in the underlying data.

\paragraph{Canary leakage.}
Canaries are planted strings inserted into synthetic conversations. We define a leak to occur when a canary string appears verbatim in a published summary. We report both a per-canary leak rate and a cluster-level leak rate (fraction of canary-containing clusters whose summaries contain at least one canary). This definition is intentionally strict, focusing on unambiguous exposure.

\paragraph{PII indicator hits.}
We additionally track coarse PII-like indicators in summaries using regular expressions for email addresses, phone numbers, and ZIP-code-like strings. This does not guarantee that all sensitive information is detected, but it provides an interpretable secondary signal.

\section{Formalization and leakage metrics}

\subsection{Problem setup}
Let $\mathcal{C} = \{c_1, c_2, \ldots, c_N\}$ be a corpus of $N$ conversations, where each conversation $c_i$ is a sequence of tokens. A clustering function $f_{\text{cluster}}: \mathcal{C} \to \{1, \ldots, K\}$ assigns each conversation to one of $K$ clusters. For cluster $j$, let $\mathcal{C}_j = \{c_i : f_{\text{cluster}}(c_i) = j\}$ denote the set of conversations assigned to cluster $j$.

A summarization function $f_{\text{summ}}: \mathcal{C}_j \to S_j$ produces a textual summary $S_j$ for each cluster. The published output is the set of summaries $\mathcal{S} = \{S_1, S_2, \ldots, S_K\}$.

\subsection{Canary injection and detection}
We define a canary set $\mathcal{W} = \{w_1, w_2, \ldots, w_M\}$ containing $M$ unique planted strings. Each canary $w_m$ is injected into a subset of conversations with probability $p_{\text{canary}}$. In our experiments, $p_{\text{canary}} = 0.60$.

For a given conversation $c_i$, let $\text{canaries}(c_i) \subseteq \mathcal{W}$ denote the set of canaries present in $c_i$. For a cluster $j$, the cluster-level canary set is:
\[
\mathcal{W}_j = \bigcup_{c_i \in \mathcal{C}_j} \text{canaries}(c_i)
\]

A canary $w_m$ leaks in cluster $j$ if $w_m$ appears verbatim as a substring in the published summary $S_j$. Formally:
\[
\text{leaked}(w_m, S_j) = \mathbb{1}[w_m \in \text{substrings}(S_j)]
\]

\subsection{Leakage metrics}
We define two complementary metrics:

\paragraph{Per-canary leak rate.}
The fraction of canary instances that leak. For each conversation $c_i$ containing canaries, we check if those canaries appear in the cluster summary. Formally, let $j_i = f_{\text{cluster}}(c_i)$ be the cluster containing conversation $c_i$. Then:
\[
\text{LR}_{\text{canary}} = \frac{\sum_{c_i \in \mathcal{C}_{\mathrm{pub}}} \sum_{w_m \in \text{canaries}(c_i)} \text{leaked}(w_m, S_{j_i})}{\sum_{c_i \in \mathcal{C}_{\mathrm{pub}}} |\text{canaries}(c_i)|}
\]
The numerator counts how many conversation--canary pairs $(c_i, w_m)$ have $w_m$ appearing in the cluster summary. In configurations that suppress some clusters (for example via k-min thresholding), leakage is evaluated over the conversations that belong to published clusters. Accordingly, the denominator is the total number of canary instances among conversations in published clusters for that configuration, which can differ across runs. In our extractive run: 51 leaked instances out of 1835 total instances, giving 0.0278.

\paragraph{Cluster-level leak rate.}
The fraction of canary-containing clusters that leak at least one canary:
\[
\text{LR}_{\text{cluster}} = \frac{1}{|\{j : |\mathcal{W}_j| > 0\}|} \sum_{j : |\mathcal{W}_j| > 0} \mathbb{1}\left[\exists w_m \in \mathcal{W}_j : \text{leaked}(w_m, S_j)\right]
\]
This measures whether clusters with sensitive content are likely to expose that content.

\subsection{Defense mechanisms}
We evaluate two defenses that modify the published summary set:

\paragraph{k-min publication threshold.}
Only publish summaries for clusters satisfying $|\mathcal{C}_j| \geq k_{\text{min}}$. Let:
\[
\mathcal{S}_{\text{pub}} = \{S_j : |\mathcal{C}_j| \geq k_{\text{min}}\}
\]
In our experiments, $k_{\text{min}} = 25$.

\paragraph{Redaction function.}
Apply a redaction function $f_{\text{redact}}: S_j \to S'_j$ that removes or masks substrings matching PII patterns. Let $\mathcal{P}$ be a set of regular expression patterns (emails, phones, ZIPs, canary-like strings). The redacted summary is:
\[
S'_j = f_{\text{redact}}(S_j, \mathcal{P})
\]
where all matches to patterns in $\mathcal{P}$ are replaced with placeholder tokens.

The defended configuration applies both mechanisms: $(k_{\text{min}}, f_{\text{redact}})$ to produce the final published set.

\subsection{Why this matters}
The key insight is that even when individual conversations are never published, aggregate summaries can leak identifying strings if the summarization function $f_{\text{summ}}$ is extractive. If $f_{\text{summ}}$ copies text directly from member conversations, then for any canary $w_m \in \mathcal{W}_j$, there exists a direct path for $w_m$ to appear in $S_j$. This is fundamentally different from statistical aggregates (counts, averages) which cannot leak verbatim strings.

The formalization makes clear that verbatim leakage is enabled by the summarization function rather than by clustering itself. For extractive summaries, leakage becomes likely when the exemplar selection includes a canary-bearing span, so high leakage should be treated as an expected outcome rather than a surprising corner case.

\begin{figure}[t]
\centering
\small
\begin{tcolorbox}[colback=blue!5, colframe=blue!60, boxrule=0.5pt, arc=2pt, title=\textbf{Example: How Extractive Summarization Leaks Canaries}]

\textbf{Input: 3 conversations in a cluster about Python}

\textcolor{gray}{Conversation 1:} ``Help me debug this Python code. Contact: alex.patel.42157@example.com''

\textcolor{gray}{Conversation 2:} ``What's the best way to learn Python as a beginner?''

\textcolor{gray}{Conversation 3:} ``My Python script crashes when parsing JSON files.''

\vspace{2mm}
\textbf{Extractive Summary} (LEAKS):

``Representative examples: (1) Help me debug this Python code. Contact: alex.patel.42157@example.com, (2) What's the best way to learn Python as a beginner?''

\textcolor{red}{$\rightarrow$ Canary \texttt{alex.patel.42157@example.com} leaked verbatim}

\vspace{2mm}
\textbf{Non-Extractive Summary} (NO LEAK):

``Topics: Python debugging, code errors, learning resources, JSON parsing''

\textcolor{green!60!black}{$\rightarrow$ No canary leaked, only high-level keywords}

\end{tcolorbox}
\caption{Extractive summarization directly quotes conversation text, creating a path for canaries to appear in published summaries. Non-extractive approaches aggregate topics without verbatim reproduction, preventing direct leakage.}
\label{fig:leak_example}
\end{figure}

\section{Benchmark and experimental setup}
\subsection{Synthetic conversation generation}
CanaryBench generates synthetic single-turn conversations across multiple topics. Each conversation includes a task prompt and may include:
\begin{itemize}
  \item \textbf{Canary strings}, injected with probability 0.60 in our main run. Canaries include synthetic emails (pattern \texttt{alex.patel.<5 digits>@example.com}), synthetic phone numbers (pattern \texttt{+1-415-555-<4 digits>}), synthetic addresses, and short unique phrases.
  \item \textbf{Additional PII-like strings}, injected with the generator default probability 0.20 when not overridden. These are also synthetic and are used to probe detector behavior.
\end{itemize}

The generator and patterns are designed to avoid real identities while preserving the structure of common identifiers.

\subsection{Embedding and clustering}
Conversations are embedded using TF-IDF features \cite{salton1988term}. We cluster embeddings using k-means \cite{lloyd1982least,macqueen1967methods}. In the reported run with 3{,}000 conversations, the pipeline produced 54 clusters in the leaky baseline configuration, matching the run output.

\subsection{Cluster-level summarization}
We evaluate two summary strategies:
\begin{itemize}
  \item \textbf{Keyword-based summary} (non-extractive): summarizes clusters using high-weighted terms, avoiding direct quotation of conversation text.
  \item \textbf{Extractive example-based summary}: summarizes clusters by surfacing representative text from within a cluster. In our runs, this corresponds to an \texttt{example} summarizer configuration, which produced measurable leakage in the outputs reported below.
\end{itemize}
We use the extractive example-based summarizer as the primary stress test in the quantitative results section, since it intentionally models quote-like reporting. The keyword-based summarizer is included as a non-extractive reference that illustrates how leakage is avoidable when summaries do not copy raw user text.

Our purpose in using the extractive strategy is to model a realistic failure mode where summaries copy or closely track user-provided strings.

\subsection{Defenses}
We evaluate a minimal defense composed of two operational controls:
\begin{itemize}
  \item \textbf{k-min publication threshold}: only publish summaries for clusters with at least $k$ conversations (here $k=25$). This is motivated by k-anonymity-style intuitions about small groups \cite{sweeney2002kanonymity}.
  \item \textbf{Regex-based redaction}: remove matches for emails, phone numbers, ZIP-like strings, and canary-like patterns from summary text, similar in spirit to practical PII anonymization frameworks \cite{presidio_github}.
\end{itemize}

\subsection{Utility proxy}
We report the average cluster coherence metric emitted by the implementation, which measures how tightly clustered the conversations are under the embedding representation. We treat this only as a coarse proxy for topical grouping quality.

\section{Results}
Table \ref{tab:results} reports the results from the two configurations you executed on the high-canary dataset: (1) extractive example-based summaries without defenses, and (2) defended summaries with k-min = 25 and redaction enabled.

\begin{table}[t]
  \centering
  \footnotesize
  \caption{Results on 3{,}000 synthetic conversations (24 topics, canary rate 0.60). Abbreviations: Pub=Published clusters, Can=Canaries, w/=with, Coh=Coherence.}
  \label{tab:results}
  \setlength{\tabcolsep}{3pt}
  \begin{tabular}{@{}lccccccccc@{}}
    \toprule
    Configuration & Pub & Can & \shortstack{Canary\\leaks} & \shortstack{Leak\\rate} & \shortstack{Clust\\w/ can} & \shortstack{Clust\\leaking} & \shortstack{Cluster\\leak rate} & \shortstack{PII\\E/P/Z} & Coh \\
    \midrule
    Extractive example & 54 & 1835 & 51 & 0.028 & 52 & 50 & 0.962 & 17/20/27 & 0.653 \\
    Defended (k-min+redact) & 32 & 1699 & 0 & \textbf{0.000} & 32 & 0 & \textbf{0.000} & 0/0/0 & 0.662 \\
    \bottomrule
  \end{tabular}
\end{table}

\begin{figure}[t]
  \centering
  \includegraphics[width=0.85\linewidth]{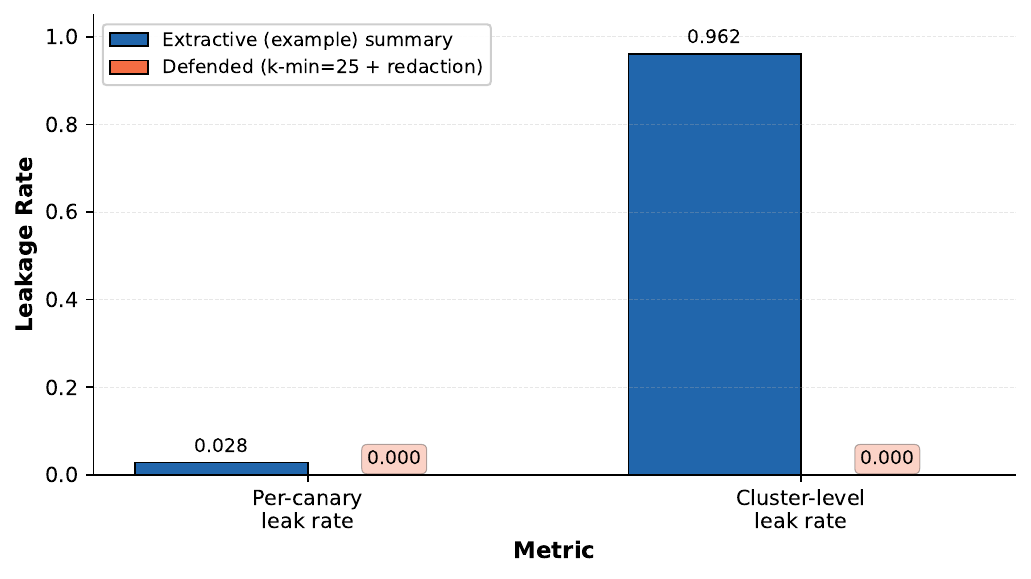}
  \caption{Leakage rates under the extractive example-based summarizer. The defended configuration reduces both per-canary and cluster-level leakage to zero in the reported runs. Zero values are annotated with highlighted boxes for clarity.}
  \label{fig:leakagerates}
\end{figure}

\begin{figure}[t]
  \centering
  \includegraphics[width=0.85\linewidth]{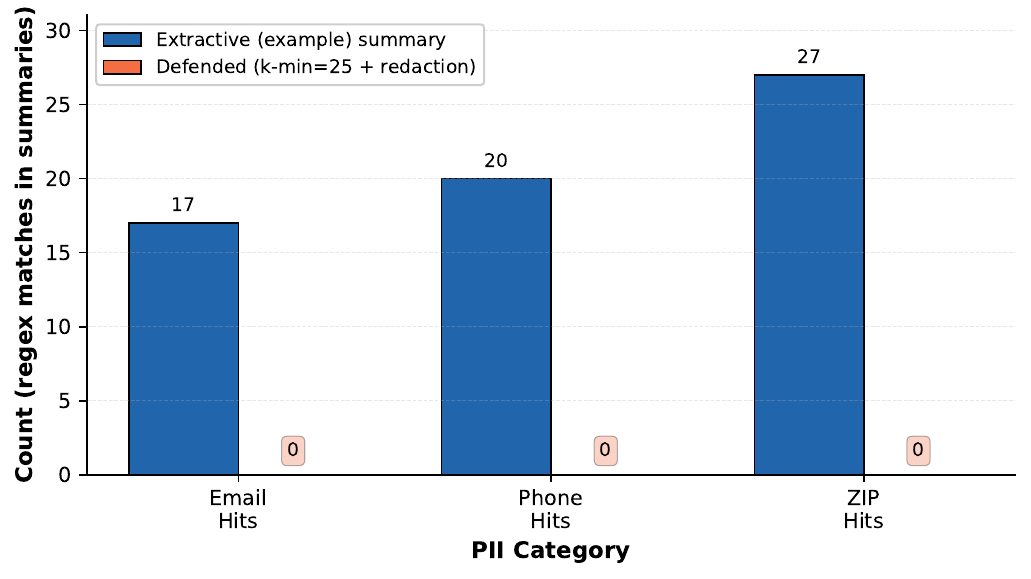}
  \caption{Regex-based PII indicator hits in published summaries. Counts drop to zero after applying redaction. Zero values are clearly annotated.}
  \label{fig:piihits}
\end{figure}

\begin{figure}[t]
  \centering
  \includegraphics[width=0.75\linewidth]{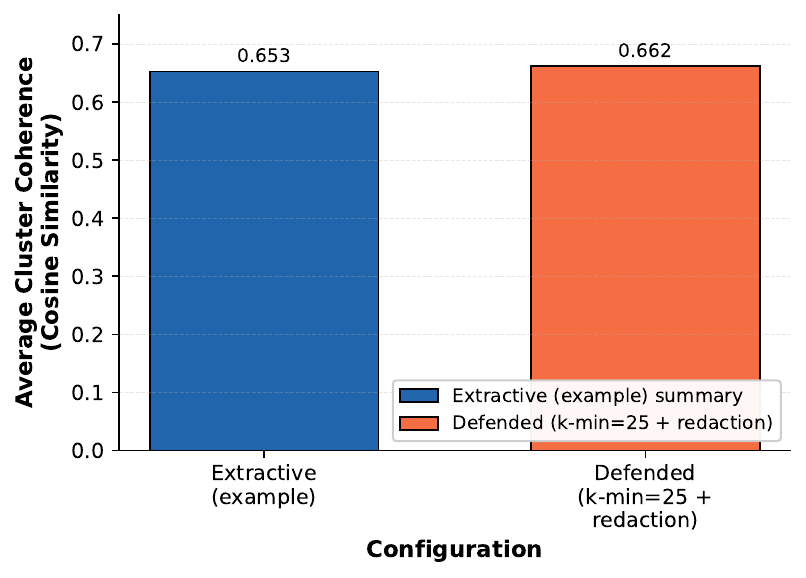}
  \caption{Utility proxy (average cluster coherence) in the reported runs. Both configurations maintain similar coherence scores, suggesting privacy defenses do not substantially degrade topical quality.}
  \label{fig:coherence}
\end{figure}

\section{Discussion}
\paragraph{Why extractive summaries are high risk.}
In the extractive configuration, the summary text directly includes content drawn from a member conversation. When identifiers appear in any conversation within a cluster, there is a direct path for them to reappear in the published summary, yielding measurable verbatim leakage. This is consistent with broader observations that verbatim strings are a particularly brittle privacy surface \cite{carlini2021extracting}.

The measured cluster-level leakage rate of 96.2\% demonstrates that extractive summarization is unsuitable for privacy-sensitive conversational analytics. Even a single leaked canary could enable re-identification if an adversary has auxiliary information about the user's topics or writing patterns. In production systems, this could expose users to targeted phishing, doxing, or harassment based on the topics they discuss with LLM assistants.

\paragraph{Role of k-min thresholding.}
The k-min threshold reduces the number of published clusters (54 to 32 in the reported runs), thereby suppressing smaller clusters that are more likely to correspond to narrow or unique behaviors. This aligns with classical intuitions that small groups increase re-identification risk \cite{sweeney2002kanonymity}. 

Concretely, a cluster with only 5 conversations represents at most 5 individuals. If one conversation contains a unique identifier and the cluster summary is extractive, that identifier becomes associated with a set of at most 4 other people, dramatically increasing re-identification risk. Setting k-min = 25 ensures that any published summary aggregates at least 25 conversations, so any identifier-like string that still appears in a published summary is co-mingled with at least 24 other conversations. This is a conversation-level aggregation guarantee, not necessarily a distinct-user guarantee, and it should be interpreted as an operational risk-reduction heuristic rather than a formal privacy bound. However, k-anonymity alone does not guarantee privacy when quasi-identifiers can be combined or when the published text is highly specific.

\paragraph{Role of redaction.}
Regex-based redaction directly targets identifier-like strings in the summary text. In the reported defended run, redaction eliminates both canary matches and regex-based PII indicator hits. The effectiveness stems from the structured, predictable format of common identifiers like emails, phone numbers, and ZIP codes.

However, regex-based approaches have known limitations: they miss paraphrased PII, descriptive identifiers ("the developer from Palo Alto who posted about their startup on January 15th"), and domain-specific identifiers. Redaction should be viewed as a necessary baseline defense rather than a complete solution \cite{presidio_docs}. Production systems should consider combining regex-based redaction with learned PII recognizers and manual review of high-risk cluster summaries.

\paragraph{Privacy and utility tradeoffs.}
The defended run retains a similar coherence score (0.662 vs 0.653). This does not prove that summaries are equally useful to humans, but it suggests that these defenses can mitigate leakage without obviously collapsing topical structure. The 41\% reduction in published clusters (54 to 32) represents a genuine utility cost: smaller, potentially interesting clusters are suppressed to maintain privacy.

A natural extension is to run ablations (k-min only, redaction only) and sweeps over $k \in \{10, 25, 50, 100\}$ to better characterize the privacy-utility frontier. Organizations should set k-min based on their risk tolerance, with higher values for sensitive domains (mental health, legal advice) and lower values for less sensitive topics.

\paragraph{Generalization to production systems.}
While CanaryBench uses synthetic data and controlled canary injection, the failure mode it reveals applies directly to real systems: verbatim leakage through extractive summaries. Any cluster-level analytics pipeline that includes extractive text or examples in published outputs is vulnerable to the same leakage pattern. The 96.2\% cluster-level leakage rate in the undefended configuration shows that without explicit safeguards, such systems will routinely leak identifiers.

Production deployments should implement at minimum: (1) non-extractive summarization, (2) k-min thresholding with $k \geq 25$, (3) regex and learned PII redaction, and (4) manual review of summaries before publication. For especially sensitive domains, organizations should consider not publishing cluster summaries at all, instead using aggregate statistics that cannot leak individual identifiers.

\section{Societal impacts}
Aggregate conversation analytics can help with safety monitoring and governance, but can also create harm if derived artifacts expose sensitive information. The risks are especially acute for vulnerable populations and sensitive topics.

\paragraph{Re-identification and targeted harm.}
Leakage of identifying strings enables re-identification attacks. For example, if a cluster summary leaks the email "alex.patel.42157@example.com" alongside the topic "anxiety medication side effects," an adversary who knows this email address can now associate that individual with mental health concerns. This creates risks of:
\begin{itemize}
  \item \textbf{Targeted phishing and social engineering}: attackers can craft convincing phishing emails referencing the leaked topics
  \item \textbf{Discrimination}: employers, insurers, or landlords who gain access to leaked summaries could discriminate based on health status, political views, or other protected characteristics
  \item \textbf{Harassment and doxing}: leaked identifiers combined with controversial topics enable targeted harassment campaigns
\end{itemize}

\paragraph{Chilling effects on help-seeking behavior.}
Even absent explicit identifiers, publishing narrow or stigmatizing cluster summaries may create chilling effects. Users may avoid asking LLM assistants for help with:
\begin{itemize}
  \item Mental health crises, substance abuse, or domestic violence
  \item LGBTQ+ identity questions in hostile jurisdictions
  \item Whistleblowing or reporting workplace misconduct
  \item Legal questions about immigration, discrimination, or family law
\end{itemize}
The fear that one's sensitive questions could appear in published analytics (even when aggregated) can deter help-seeking behavior and reduce the social value of LLM assistants as a resource for vulnerable populations.

\paragraph{Balancing transparency and privacy.}
Conversely, overly restrictive analytics can reduce transparency and hinder safety monitoring. Organizations deploying LLM systems have legitimate needs to understand usage patterns, identify misuse, and improve safety. The challenge is to enable useful analytics while protecting individual privacy.

Our results support conservative defaults for any system that publishes aggregate summaries derived from conversations:
\begin{itemize}
  \item \textbf{Never use extractive summarization} for public or semi-public analytics releases
  \item \textbf{Enforce minimum aggregation thresholds} (we use $k=25$ in our experiments; more sensitive domains may warrant larger $k$ values selected via policy and validation)
  \item \textbf{Apply multi-layer redaction}: regex patterns, learned PII recognizers, and manual review
  \item \textbf{Treat analytics outputs as sensitive artifacts} with appropriate access controls, audit logs, and retention policies
  \item \textbf{Consider differential privacy mechanisms} for quantitative aggregate statistics
  \item \textbf{Provide user controls} allowing individuals to opt out of analytics or delete their conversation history
\end{itemize}

Organizations should conduct privacy impact assessments before deploying cluster-level analytics and establish clear policies on what can be published, who can access analytics, and how long data is retained.

\section{Limitations}
This study uses synthetic conversations rather than real user data. While synthetic canaries provide clean ground truth for leakage measurement, they may not capture the full distribution of identifiers and context in real deployments. Our leakage definition is strict (verbatim matching) and therefore does not capture semantic leakage or attribute inference. Our PII indicator detection uses regular expressions and will miss some sensitive content. Finally, the coherence metric is only a proxy for topical grouping quality and does not substitute for human evaluation.

\section{Reproducibility}
We will release code and scripts used to produce the reported runs at \url{https://github.com/researchaudio/canarybench}.

The following commands reproduce the reported configurations and are included verbatim for reproducibility:

\begin{lstlisting}[basicstyle=\ttfamily\small]
python run_experiment.py generate --out data/convos_highleak.jsonl \
  --n 3000 --topics 24 --canary_rate 0.60

python run_experiment.py run --data data/convos_highleak.jsonl \
  --out runs/leaky2 --embed tfidf --cluster kmeans --summarizer example

python run_experiment.py run --data data/convos_highleak.jsonl \
  --out runs/leaky2_defended --embed tfidf --cluster kmeans \
  --summarizer example --k_min 25 --redact_pii
\end{lstlisting}

\section{Conclusion}
CanaryBench provides a simple, reproducible stress test for privacy leakage in cluster-level summaries derived from conversational data. Our key findings are:

\begin{itemize}
  \item \textbf{Extractive summaries leak at scale}: With extractive example-based summarization, we observe canary leakage in 50 of 52 canary-containing clusters (96.2\% cluster-level leakage rate), demonstrating that extractive approaches are fundamentally unsuitable for privacy-sensitive analytics.
  
  \item \textbf{Minimal defenses are highly effective}: A simple defense combining k-min thresholding (k=25) with regex-based redaction reduces measured leakage to zero while preserving topical coherence, showing that straightforward operational safeguards can substantially mitigate privacy risks.
  
  \item \textbf{Privacy-utility tradeoffs are manageable}: The defended configuration maintains similar coherence (0.662 vs 0.653) despite suppressing 41\% of clusters, suggesting that privacy protections need not collapse analytical utility for aggregate topic understanding.
\end{itemize}

\paragraph{Recommendations for practitioners.}
Organizations deploying cluster-level analytics over user conversations should:
\begin{enumerate}
  \item Never use extractive summarization for published analytics
  \item Implement k-min thresholding with $k \geq 25$ (higher for sensitive domains)
  \item Apply regex-based and learned PII redaction to all published text
  \item Manually review cluster summaries before publication
  \item Conduct privacy impact assessments and establish data retention policies
  \item Provide users with opt-out mechanisms and data deletion controls
\end{enumerate}

\paragraph{Future work.}
CanaryBench's focus on exact string matching provides a lower bound on leakage risk. Future work should explore: (1) semantic leakage through paraphrased or aggregate information, (2) re-identification attacks using quasi-identifiers and auxiliary information, (3) privacy-utility tradeoffs across different values of k and alternative anonymization mechanisms, (4) evaluation on real conversational data with ground-truth PII labels, and (5) integration with differential privacy guarantees for stronger formal privacy protection.

We hope CanaryBench encourages routine, reproducible leakage evaluation as a standard practice for any system that publishes aggregate text derived from user conversations. The benchmark is designed to be simple enough to serve as a continuous integration check: if cluster summaries leak canaries, they should not be published.

\appendix
\section{Minimal redaction and leakage check used in the starter kit}
To make the defense and evaluation surface explicit, we include the exact regular expressions and leakage check used in the accompanying starter implementation. These snippets are reproduced from the redaction and evaluation modules.

\subsection{Regex-based redaction}
\begin{lstlisting}[basicstyle=\ttfamily\small]
import re
from typing import Dict, List, Tuple

EMAIL_RE = re.compile(r"\b[\w.+'-]+@[\w.-]+\.[A-Za-z]{2,}\b")
PHONE_RE = re.compile(r"(\+?\d{1,3}[\s-]?)?(\(?\d{3}\)?[\s-]?)\d{3}[\s-]?\d{4}\b")
ZIP_RE = re.compile(r"\b\d{5}(?:-\d{4})?\b")

def redact_pii(text: str) -> str:
    # Light-weight, regex-based redaction for the starter kit.
    text = EMAIL_RE.sub("[REDACTED_EMAIL]", text)
    text = PHONE_RE.sub("[REDACTED_PHONE]", text)
    # addresses are hard; we at least remove zip-like tokens
    text = ZIP_RE.sub("[REDACTED_ZIP]", text)
    return text
\end{lstlisting}

\subsection{Exact canary leakage check}
\begin{lstlisting}[basicstyle=\ttfamily\small]
def exact_canary_leaks(summary: str, canaries: List[Dict]) -> Tuple[int,int]:
    # returns (num_leaked, num_total)
    total = len(canaries)
    leaked = 0
    for c in canaries:
        if c["value"] in summary:
            leaked += 1
    return leaked, total
\end{lstlisting}

\end{document}